\documentclass{aa}
\usepackage{graphics}
\newcommand{\cmtwo}{cm$^{-2}$}  
\newcommand{\cmthree}{cm$^{-3}$}

\newcommand{\kms}{km\,s$^{-1}$}       
\newcommand{\vlsr}{$v_{\rm LSR}$}        

\newcommand{\um}{$\mu$m}                                 
\newcommand{\molh}{H$_{2}$}                              

\newcommand{\water}{H$_{2}$O}

\newcommand{\lapprox}{$\stackrel {<}{_{\sim}}$}
\newcommand{\about}{$\sim$}                       
\newcommand{\powten}[1]{10$^{#1}$}
\newcommand{\amin}{$^{\prime}$}                   
\newcommand{\asec}{$^{\prime \prime}$}

\newcommand{\radot}[4]{\mbox{#1$^{\rm h}$#2$^{\rm m}$#3$\stackrel{\rm s}
{_{\bf\cdot}}$#4}}  
\newcommand{\decdms}[3]{\mbox{#1$^{\circ}$#2$^{\prime}$#3$^{\prime \prime}$}}

\begin{document}

   \title{First NH$_3$ detection of the Orion Bar\thanks{Based 
  on observations with Odin, a Swedish-led satellite project funded jointly by 
  the Swedish National Space Board (SNSB), the Canadian Space Agency (CSA), 
  the National Technology Agency of Finland (Tekes)  and Centre National
  d'Etude Spatiale (CNES). The Swedish Space Corporation has been the industrial 
  prime contractor.}
  }


   \author{     B.\,Larsson\inst{1}    	  \and
		R.\,Liseau\inst{1}     	  \and
            	P.\,Bergman\inst{2}    	  \and
		P.\,Bernath\inst{3}       \and
            	J.H.\,Black\inst{2}    	  \and 
		R.S.\,Booth\inst{2}       \and
	    	V.\,Buat\inst{4}    	  \and
		C.L.\,Curry\inst{3}    	  \and
		P.\,Encrenaz\inst{5}      \and 		 
		E.\,Falgarone\inst{6}     \and
                P.\,Feldman\inst{7}       \and 
		M.\,Fich\inst{3}    	  \and
                H.G.\,Flor\'{e}n\inst{1}  \and
		U.\,Frisk\inst{8}  	  \and  
		M.\,Gerin\inst{6}    	  \and
		E.M.\,Gregersen\inst{9}   \and 
		J.\,Harju\inst{10}    	  \and		
                T.\,Hasegawa\inst{11}     \and
                L.E.B.\,Johansson\inst{2} \and
		S.\,Kwok\inst{11}    	  \and
                A.\,Lecacheux\inst{12}    \and
		T.\,Liljestr\"om\inst{13} \and
                K.\,Mattila\inst{10}      \and
		G.F.\,Mitchell\inst{14}   \and
		L.H.\,Nordh\inst{15}      \and
		M.\,Olberg\inst{2}    	  \and
		G.\,Olofsson\inst{1}      \and
		L.\,Pagani\inst{5}	  \and
		R.\,Plume\inst{11}	  \and
		I.\,Ristorcelli\inst{16}  \and 
                Aa.\,Sandqvist\inst{1}    \and
                F.v.\,Sch\'eele\inst{8}   \and
		N.F.H.\,Tothill\inst{14}  \and
                K.\,Volk\inst{9}          \and
		C.D.\,Wilson\inst{9} 	  \and 
		\AA.\,Hjalmarson\inst{2}
	}

   \offprints{B. Larsson}

   \institute{ Stockholm Observatory, SCFAB, Roslagstullsbacken 21, SE-106 91 Stockholm, Sweden \\
   	\email{bem@astro.su.se}
    \and  
         Onsala Space Observatory, SE-439 92, Onsala, Sweden         
    \and         
         Department of Physics, University of Waterloo, Waterloo, ON N2L 3G1, Canada
    \and      
         Laboratoire d'Astronomie Spatiale, BP 8, 13376 Marseille CEDEX 12, France
    \and       
    	LERMA \& FRE 2460 du CNRS, Observatoire de Paris, 61, Av. de l'Observatoire, 75014 Paris, France
    \and
    	LERMA \& FRE 2460 du CNRS, Ecole Normale Sup\'erieure, 24 rue Lhomond, 75005 Paris, France
    \and
	Herzberg Institute of Astrophysics, 5071 West Saanich Road, Victoria, BC, V9E 2E7, Canada
    \and       
    	Swedish Space Corporation, P O Box 4207, SE-171 04 Solna, Sweden   
    \and    	
    	Department of Physics and Astronomy, McMaster University, Hamilton, ON, L8S 4M1, Canada
    \and    	
    	Observatory, P.O. Box 14, University of Helsinki, 00014 Helsinki, Finland
    \and    	
    	Department of Physics and Astronomy, University of Calgary, Calgary, ABT 2N 1N4, Canada
    \and
	LESIA, Observatoire de Paris, Section de Meudon, 5, Place Jules Janssen, 92195 MEUDON CEDEX, France
    \and   
    	Mets\"ahovi Radio Observatory, Helsinki University of Technology, Otakaari 5A, FIN-02150 Espoo, Finland    
    \and      
    	Department of Astronomy and Physics, Saint Mary's University, Halifax, NS, B3H 3C3, Canada
    \and      
        Swedish National Space Board, Box 4006, SE-171 04 Solna, Sweden
    \and    
    	CESR, 9 Avenue du Colonel Roche, B.P. 4346, F-31029 Toulouse, France
    }

\date{Received date: \hspace{5cm}Accepted date:}

   \abstract{Odin has successfully observed three regions in the Orion\,A cloud, i.e. Ori\,KL, Ori\,S and the Orion Bar, 
   	in the 572.5\,GHz rotational ground state line of ammonia, ortho-NH$_3$ 
	$(J,K) = (1,0) \rightarrow (0,0)$, and 
        the result for the Orion Bar represents the first detection in an ammonia line. Several velocity components are
        present in the data. Specifically, the observed line profile from the Orion Bar can be decomposed into two components, 
        which are in agreement with observations in high-$J$ CO lines by \cite{wilson01}. Using the source model for the Orion Bar
        by these authors, our Odin observation implies a total ammonia abundance of ${\rm NH}_3/{\rm H}_2 = 5\times 10^{-9}$. 
         \keywords{ ISM: individual objects: Orion A  -- clouds -- molecules -- abundances --
                   Stars: formation} 
               }
   \maketitle

%

\section{Introduction}

The central parts of the Orion\,A cloud contain several different source regions, among which one can identify
Orion\,KL, Orion\,S and the Orion Bar (see, e.g., Fig.\,3 of \cite{wilson01}). On angular scales of
two arcminutes (the Odin beam size, Sect.\,2), the Ori\,KL region, in particular, harbours a highly complex 
source structure, apparently with several components along the line of sight. The region has been 
reviewed by \cite{genzel89} and more recent results and references can be found in the papers of 
\cite{wilson01} and of \cite{wisemanho98}. The latter authors presented VLA-maps in two inversion lines of 
para-ammonia, showing the emission to arise from finger-like structures. However, as in previous observations 
(e.g. \cite{ho79}), the Orion Bar was not detected.

The structure of the NH$_3$ molecule and its versatility as an astrophysical tool have been discussed by 
\cite{hoandtownes83}. These authors also provide an energy level diagram. The rotational ground state 
line of ammonia, NH$_3$ 
$(J,K) = (1,0) \rightarrow (0,0)$, has a wavelength of 524\,\um\ and an upper state
energy of 28\,K above ground. For the
$(1,1)$ and $(2,2)$ inversion lines 
at 1.3\,cm, the corresponding values are 24\,K and 65\,K, respectively. The radiative lifetime of the $(1,0) \rightarrow (0,0)$
transition is shorter by orders of magnitude than those of the inversion lines and the line profile could therefore
potentially be probing regions of very different excitation conditions (cf. Table\,2 of Liseau et al.; this volume). 
In particular for dynamical studies, the NH$_3$ line is expected to complement or, in regions of undetectably low 
water vapour abundance, to substitute for the ground state line of ortho-\water\,($1_{10}-1_{01}$), since these resonance 
lines share the property of being `effectively transparent' up to very high optical depths, as they exhibit an 
essentially linear growth of the intensity (for $\tau$ of several hundred; see, e.g., \cite{snell00} and \cite{liseau01}).

Using the Kuiper Airborne Observatory (KAO), \cite{keene83} observed the NH$_3$\,
$(1,0) \rightarrow (0,0)$ line
toward Ori\,KL and made a five point cross with 1\amin\ spacing. Their observations will be compared 
to those presented in this {\it Letter}, which have been obtained with Odin (Sect.\,2), a spaceborne submillimeter 
telescope (Frisk et al., Hjalmarson et al. and Nordh et al., this volume). The reductions of
these data, which required some special and careful treatment, are described in detail in Sect.\,3
and the results are presented in Sect.\,4. The concluding discussion (Sect.\,5) of this {\it Letter} will focus on the 
NH$_3$ emission from the Orion Bar, as this provides an entirely new piece of information for the complex Orion\,A region.

\section{The Odin observations}
 
The NH$_3$ observations of Orion A were made on September 27 to 29, 2001, simultaneously with observations of the 
H$_2^{18}$O\,($1_{10}-1_{01}$) and O$_2$\,($1_1-1_0$) lines (Olofsson et al. and Pagani et al.; this volume).
Aiming at Ori\,KL, the coordinates of the (0,\,0) position were RA\,=\,\radot{5}{35}{15}{1} and Dec\,=\,\decdms{$-5$}{22}{12} (J2000).
The absolute pointing of the Odin telescope is presently known to within 30\asec. At 572.5\,GHz the half power beam width is 2\amin.
In addition, two positions with the nominal relative offsets in arcmin (0,\,$-2$; Ori\,S) and ($+2$,\,$-2$; Orion Bar) were observed.
During the observations, the relative pointing accuracy (rms) was
($\Delta {\rm RA},\,\Delta {\rm Dec})$ \lapprox (7\asec, 14\asec)
The data were obtained in a sky-switching mode (Frisk et al., Hjalmarson et al.; this volume)
by observing a total of 4.5\,hr each on-source and on blank sky, with 10\,s 
integrations per individual scan. As back-end, a digital hybrid autocorrelator (AC) was used, with an 800\,MHz
bandwidth at the resolution of 1\,MHz (0.5\,\kms).

\section{Data reductions}

The phase-lock system for the 572\,GHz receiver is not working properly. However, it is reasonably well locked at the
tuning frequency and the telluric ozone line O$_3$ ($J_{K^-,\,K^+}= 30_{4,\,26} \rightarrow 30_{3,\,27}$) 572.9\,GHz
is frequently observed along parts of the Odin orbit, drifting across the receiver band. The line is 
sufficiently close in frequency to the ammonia line NH$_3$ 
$(J,K) = (1,0) \rightarrow (0,0)$ 572.5\,GHz
that it can be used to
establish the observing frequency scale, and hence allows us to restore the NH$_3$ data. 

\begin{figure}
  \resizebox{\hsize}{!}{\rotatebox{90}{\includegraphics{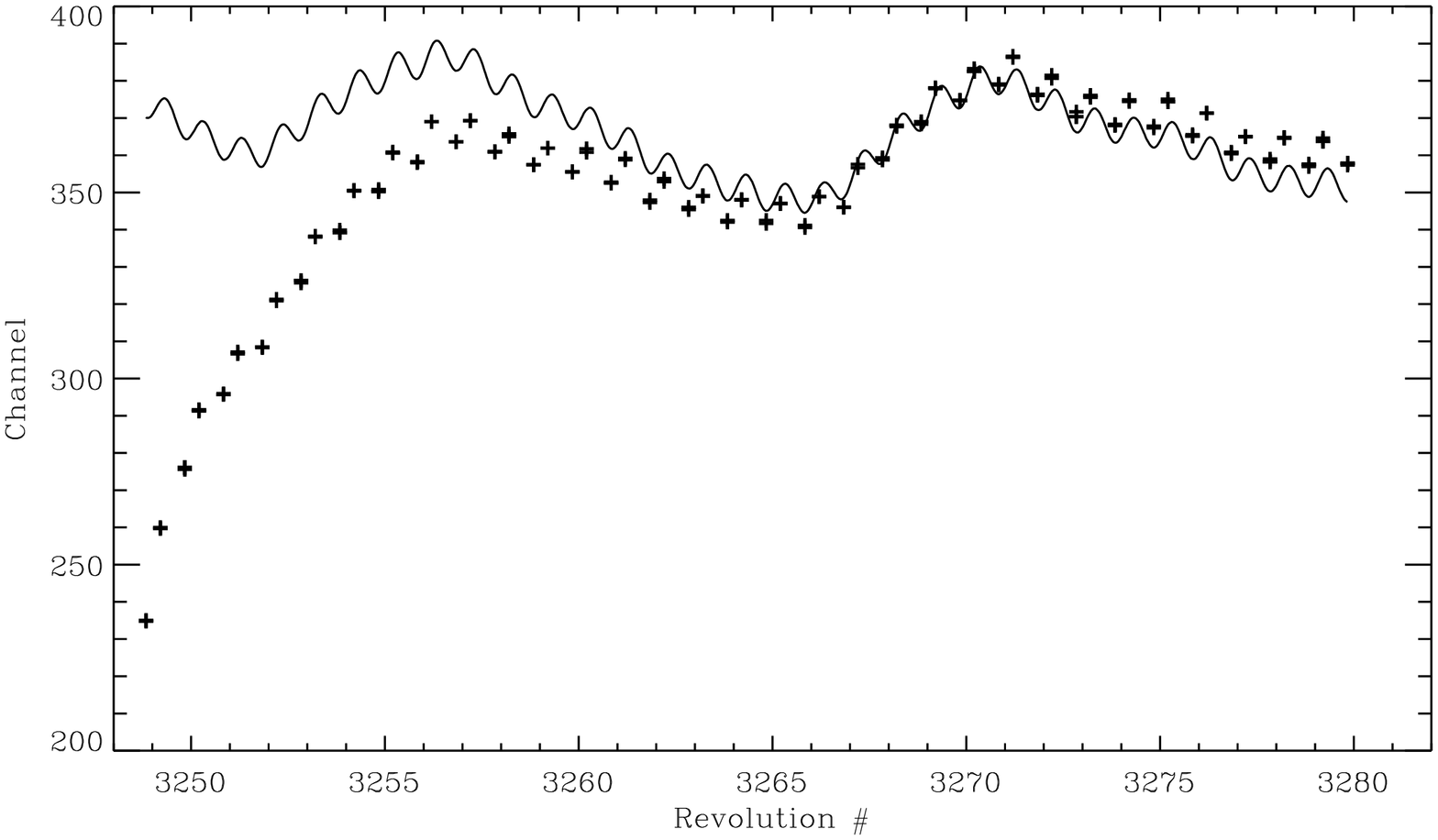}}}
  \resizebox{\hsize}{!}{\rotatebox{90}{\includegraphics{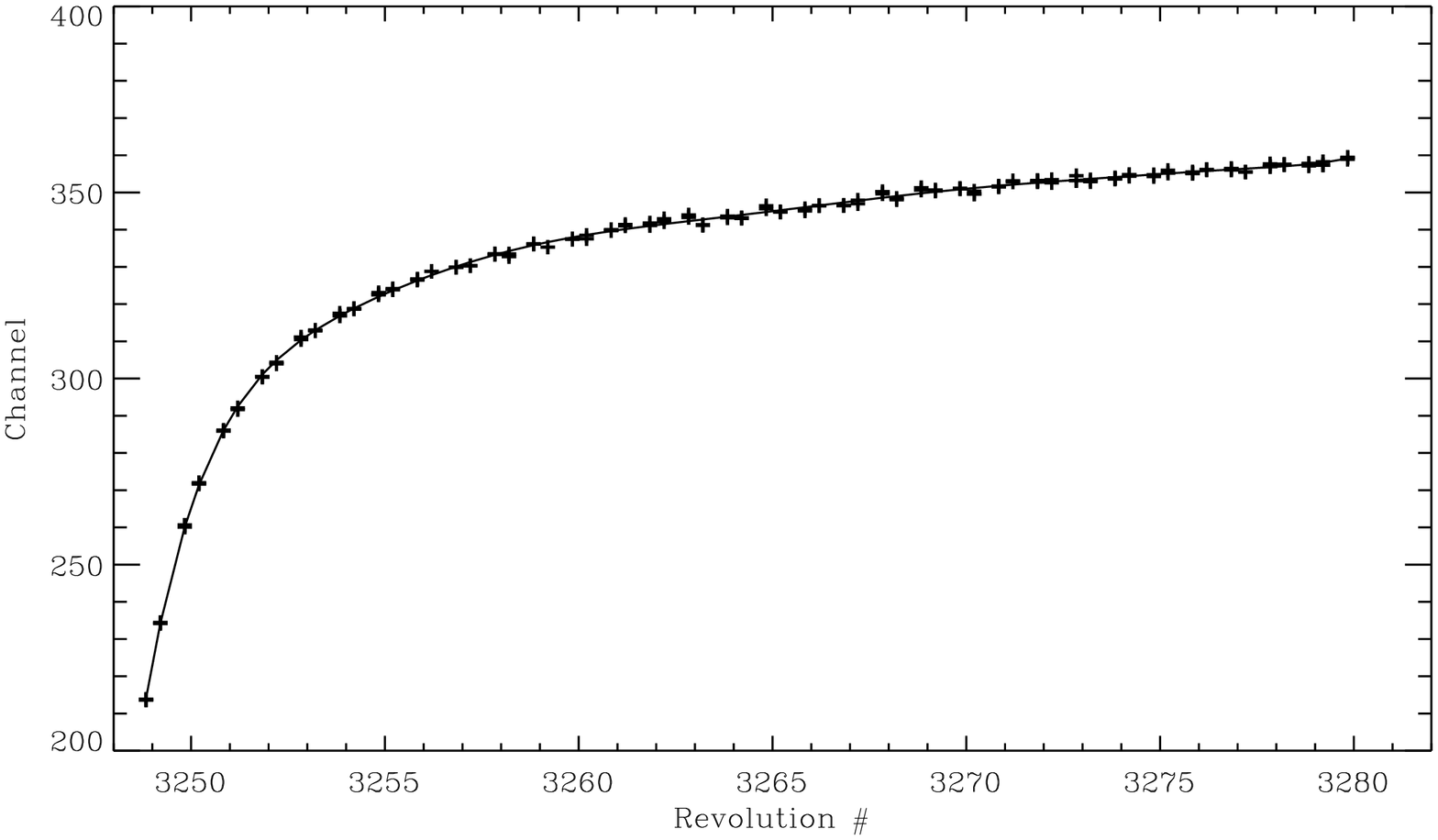}}}
  \resizebox{\hsize}{!}{\rotatebox{90}{\includegraphics{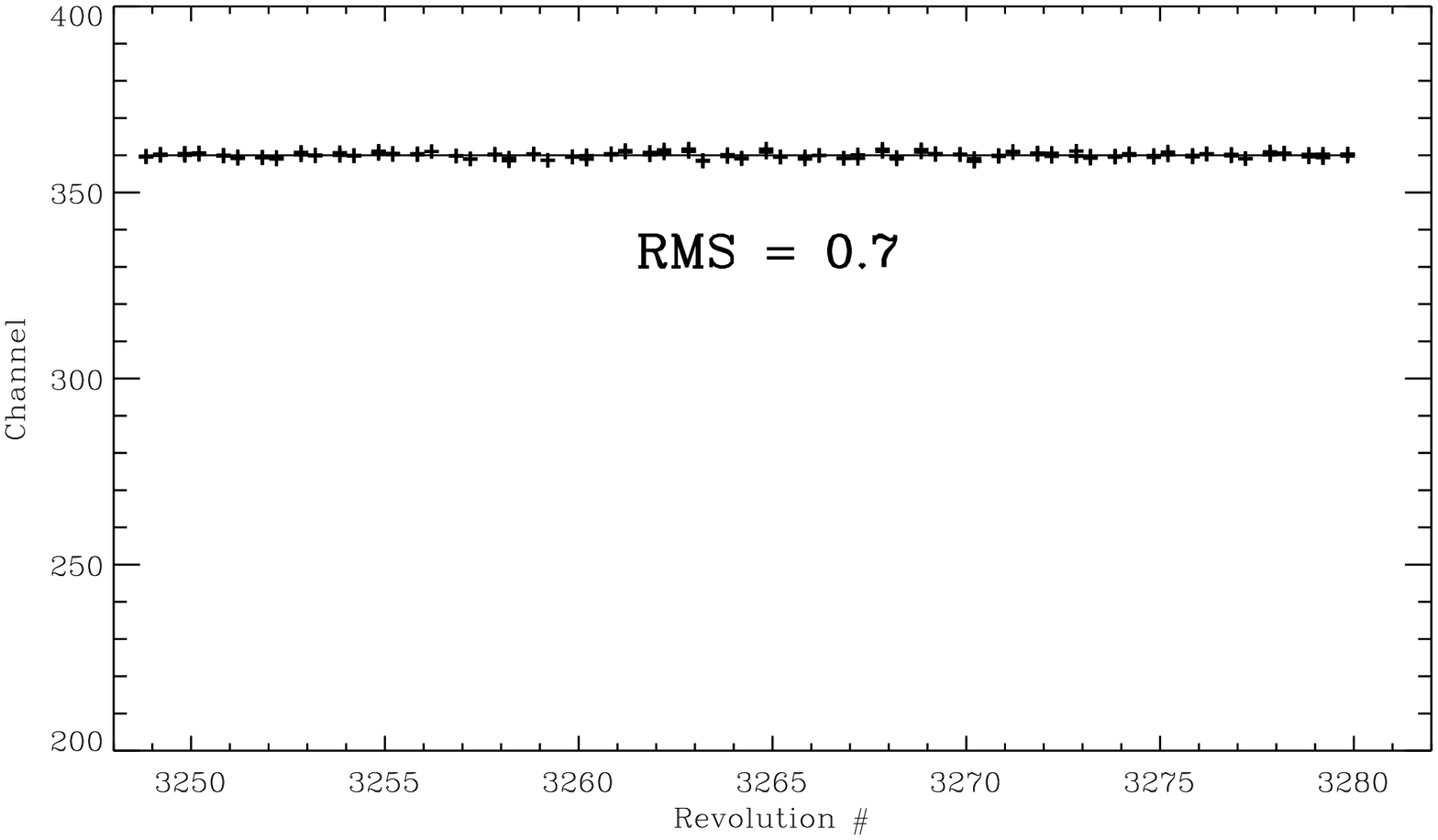}}}
  \caption{Correction for frequency drift during the observation. {\bf Upper:} The plus-signs denote the channel number of the center of the 
  telluric O$_3$ line during the Orion observations, expressed as Odin revolutions. The upper curve (solid line) depicts the behaviour of 
  the reciprocal LO-temperature.
  {\bf Middle:} The plus-signs designate the difference between the values shown above. The full-drawn curve is a polynomial fit to these
  points. This curve describes the average behaviour of the temperature measured at the main reflector.
  {\bf Lower:} The rms-spread of the residuals to the polynomial fit corresponds to 0.7 of the AC-channel width (= 0.36\,\kms).
  }
  \label{fig_atm}
\end{figure}

The drift of the O$_3$ line can be expected to correlate with the equipment temperature and, in particular, inversely with the 
temperature measured at the local oscillator (LO).
That this is indeed the case is demonstrated in the upper panel of Fig.\,\ref{fig_atm}, where the center channel for the O$_3$ line,
$C_{{\rm O}_3}$ and identified by  plus-signs, is shown as a function of the time, expressed as satellite revolution numbers. 
The solid curve in that panel depicts the relation for the reciprocal LO-temperature, $f(T_{\rm LO}) \propto 1/{T_{\rm LO}}$. 
Subtraction of the ${T_{\rm LO}}$ relation from the O$_3$ channel data points results in the swarm of plus signs in the middle 
panel of Fig.\,\ref{fig_atm}. A polynominal fit to these points is shown by the full drawn curve [$=g(\Delta)$], depicting 
the temperature behaviour measured at the telescope (main reflector). In the lower panel, the residual displays the final scatter about 
the fixed central channel for the O$_3$ line (rms = 0.7 channels = 0.36\,\kms).

The drift corrections to the Orion observations were done by applying the empirically determined relation

\begin{equation}
C_i = C_{i\,0} - [f(T_{\rm LO}) + g(\Delta) + C_{{\rm O}_3}]\,\,\,\,,
\end{equation}

where the frequencies are expressed in channel units and where the constant, $C_{{\rm O}_3}$, is derived from the ozone line observations. 
The final spectra are given in the main beam temperature scale, defined as

\begin{equation}
T_{\rm mb} = \frac{T_{\rm A}}{\eta_{\rm mb}} = \frac{1}{\eta_{\rm mb}}\, \frac{T_{\rm on} - T_{\rm sky}}{T_{\rm sky}}\, T_{\rm sys}\,\,\,,
\end{equation}

\begin{figure}
  \resizebox{\hsize}{!}{\rotatebox{90}{\includegraphics{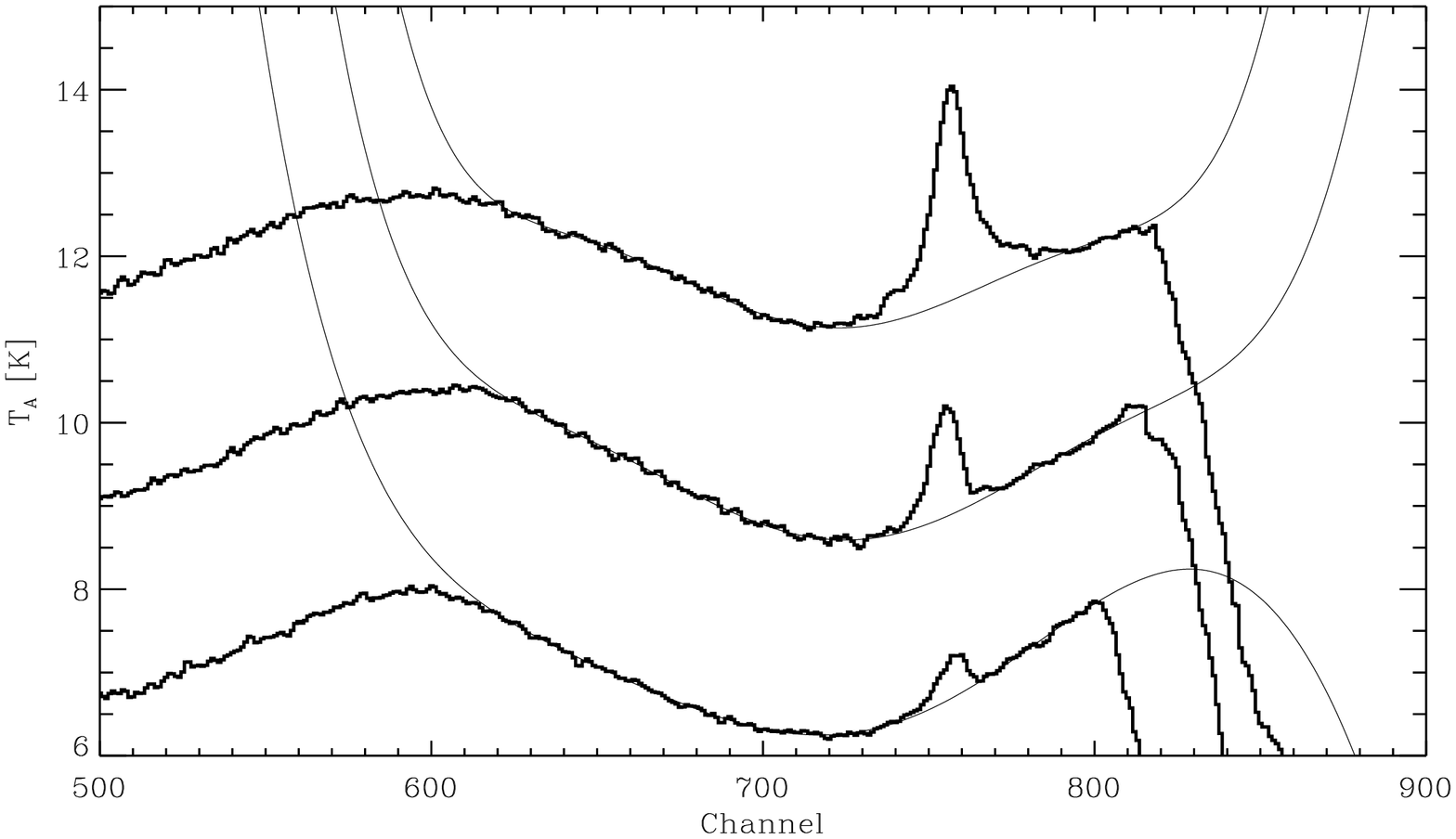}}}
  \resizebox{\hsize}{!}{\rotatebox{0}{\includegraphics{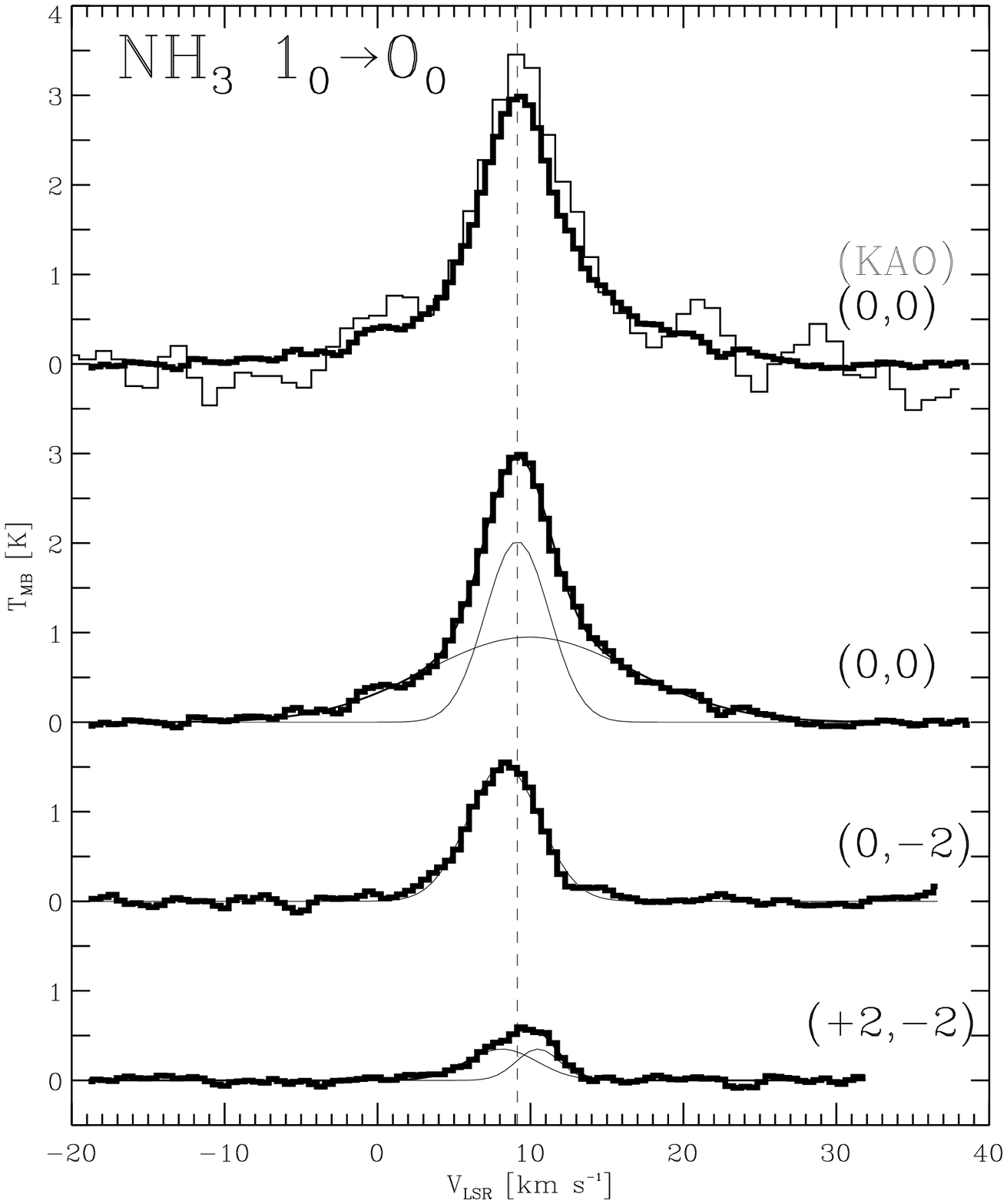}}}
  \caption{
  NH$_3$ $(J,K) = (1,0) \rightarrow (0,0)$ spectra toward Orion A.{\bf Upper:} The reduced spectra for the three positions discussed in the
  text (antenna temperature vs. correlator channel) are shown together with the adopted polynomial fits to their baselines.
  {\bf Lower:} Same as above, but with the baselines subtracted ($T_{\rm mb}$ vs. $v_{\rm LSR}$). The data are displayed as histograms. 
  Uppermost, the KAO observation (2\amin\ FWHM) by \cite{keene83} (thin line) is shown together with the Odin spectrum  toward (0, 0). The much higher
  S/N of the Odin data aside, the agreement between these two observations is strikingly good. Following below, the spectra toward the three
  positions observed with Odin are displayed, together with Gaussian fits to the line profiles as thin lines (cf. Table\,\ref{tab_fit}). 
  For reference, a vertical dashed line is drawn at \vlsr\,=\,9.1\,\kms.
  }
  \label{fig_spe}
\end{figure}

\noindent
with obvious notations (average $T_{\rm sys} = 3750$\,K). In the upper panel of
Fig.\,\ref{fig_spe}, the averaged spectral scans ($T_{\rm A}$ vs. $C$) toward the 
three observed positions in Orion are shown, together with polynomial fits to their baselines. 
Finally, the baseline subtracted spectra ($T_{\rm mb}$ vs. $v_{\rm LSR}$) are shown in the lower panel. 
The choice of polynomials is conservative, as a sinosoidal baseline fit to the remaining standing wave pattern 
would result in more extended wings in especially the (0,\,0) spectrum.

\begin{table}
\begin{flushleft}
\caption{\label{tab_fit} Gaussian fit parameters for the observed line profiles.} 
\resizebox{\hsize}{!}{  
  \begin{tabular}{c lcc l ccc} 
  \noalign{\smallskip}
  \hline
  \noalign{\smallskip}  
         & \multicolumn{3}{c}{Component 1} & &  \multicolumn{3}{c}{Component 2} \\
 \cline{2-4}  \cline{6-8} 
         
Offset   &  $T_{0,\,1}$               & $v_{{\rm LSR},\,1}$ & $\Delta v_1$ &  & $T_{0,\,2}$   & $v_{{\rm LSR},\,2}$ & $\Delta v_2$ \\
(arcmin) &  (K)                       & (\kms)              & (\kms)       &  & (K)           &  (\kms)             & (\kms) \\
  \noalign{\smallskip}
  \hline
  \noalign{\smallskip}
$(0,\,0)$     &  2.0  & 9.1 & 4.9 & & 0.9      & 9.9      & 16.0          \\
$(0,\,-2)$    &  1.5  & 8.3 & 5.5 & & $\ldots$ & $\ldots$ &  $\ldots$     \\
$(+2,\,-2)$   &  0.35 & 8.1 & 5.4 & & 0.35     & 10.5     & \phantom{1}3.3  \\
  \noalign{\smallskip}
  \hline
  \noalign{\smallskip}  
  \end{tabular}
}
\end{flushleft}
\end{table}

\section{Results}

The NH$_3$
$(J,K) = (1,0) \rightarrow (0,0)$ spectra observed toward the three positions in Ori\,A are shown
in Fig.\,\ref{fig_spe} (lower panel). The agreement with the KAO data toward Ori\,KL, also shown in that figure for
comparison, is striking, although the signal-to-noise ratio of the Odin data is far better. \cite{keene83} speculated
whether there existed some low intensity broad component in their data. That this is indeed the case has now been confirmed
by Odin. In our observations, the line intensity peaks at the KL position and decreases by about
a factor of two at Ori\,S and by a factor of five toward the Orion Bar. Changes in radial velocity of the
line centers are also evident. In the figure, the results of Gaussian fits to the line profiles
are also shown and their parameters are reported in Table\,\ref{tab_fit}. Four distinct velocity systems
can be identified: (1) a \vlsr\,=\,9\,\kms\ component of width 5\,\kms\ (FWHM) is present toward
Ori\,KL, (2) a very broad feature ($\Delta v =16$\,\kms), but centered at
\vlsr\,=\,10\,\kms, is also present in Ori\,KL, (3) a blueshifted line (\vlsr\,=\,8\,\kms) of width 5.5\,\kms\ toward the outflow 
source Ori\,S and the Orion Bar and (4) a component at \vlsr\,=\,10.5\,\kms\ of width 3.3\,\kms\ from the Orion Bar, 
identical to CO\,($J$\,=\,$4-3$) (\cite{wilson01}). 
The result for the Orion Bar represents the first detection in an ammonia line.

The radial velocities found in the NH$_3$\,
$(1,0) \rightarrow (0,0)$ spectra are different from those quoted for various inversion lines
toward Ori\,KL by \cite{hermsen88a}, but are in good agreement with those observed toward all positions
in the ortho-\water\,($1_{10}-1_{01}$) 557\,GHz line during an independent observing run  (Olofsson et al.; this volume). 
The line shapes of these transitions do agree in the direction of Ori\,S and the Bar, but at the resolution of 1\,MHz, 
the NH$_3$\,572.5\,GHz line core toward Ori\,KL is highly symmetric and does not display any self-absorption like feature 
as does the \water\,557\,GHz line.

\section{Discussion and conclusions}

For the Orion Bar, no previous NH$_3$ line information, other than  upper limit results, is in existence. 
For the interpretation of this observation, we will have to rely on source models based primarily on other data. 
The observed width of the Bar ranges from about 10\asec\ to 60\asec\ for a variety of molecules 
(e.g., \cite{tielens93}, \cite{wilson01}).
Based on observations in high$-J$ CO lines, estimates of the physical parameters of the Orion Bar have been given by \cite{wilson01}. 
These authors argue that the geometry is that of a 30\asec\ wide rod in the plane of the sky 
and we assume that their model is applicable also to NH$_3$. 
The observed width of the NH$_3$ and CO lines, $\Delta v = 3.3$\,\kms, can be 
interpreted as a velocity gradient of 45\,\kms\,pc$^{-1}$. The hyperfine structure of the NH$_3$ transition (\cite{townes1955}; 
cf. Liseau et al., this volume) does not significantly contribute to the observed width of the line.
The column density of \molh\ is given by \cite{wilson01} as $N({\rm H}_2)=2.2 \times 10^{22}$\,\cmtwo. 
As kinetic gas temperature we adopt the peak value of their observed CO (7$-$6) line, $T_{\rm k}=145$\,K. 

We use a large velocity gradient code to estimate the statistical equilibrium level populations and to solve 
the equation of radiative transfer along the line of sight. We have calculated the level energies and 
Einstein-$A$ values according to \cite{poynterandkakar75}, using the dipole moment of \cite{cohenandpoynter74}. 
For the computation of the collision rates, we adopted the values provided by \cite{danby88} for collisions 
with para-\molh\ ($J=0$), in conjunction with the assumption of detailed balance for the inverse rates. 

For the source model of \cite{wilson01}, we find the abundance of ortho-ammonia relative to \molh\
in the Orion Bar to be $X_{\rm o}({\rm NH}_3) = 2.5 \times 10^{-9}$, where the NH$_3$ ($1,0)\rightarrow (0,0$)
line radiation temperature is $T_{\rm R}=1.7$\,K, corrected  for partial beam filling (\about\,0.2). 
This result for $X_{\rm o}$ remains unaltered, if the kinetic temperature were higher, e.g., $T_{\rm k}=200$\,K. 
The model is also in agreement with the upper limits on the inversion lines of
\cite{ho79} and \cite{wisemanho98}, provided the ammonia ortho-to-para ratio is not largely different from unity.
This would be expected for a gas at such elevated temperatures, since for ammonia the ortho-to-para ratio
$\rightarrow g_{\rm o}/2\,g_{\rm p} = 1$ as $T_{\rm k} \rightarrow \infty$. The high temperature ortho-to-para
ratio of 1.0 has previously been determined for Ori\,KL by \cite{morris73}. For the Orion Bar, such ratio would thus
imply a total ammonia abundance $X({\rm NH}_3) = 5 \times 10^{-9}$. The ammonia emission region is probably not much 
narrower than 10\asec\ and average densities do probably not exceed \powten{6}\,\cmthree\ by large, so that it is likely that
$X({\rm NH}_3) < 2 \times 10^{-8}$. These estimates can be compared to
the values deduced for the Ori\,KL components `hot core' ($10^{-7}-10^{-6}$, \cite{ewine93}, \cite{hermsen88b}) 
and `plateau' ($<10^{-8}$, \cite{ewine93}). 
The density of the applied model is lower than the critical density of the ($1,0) \rightarrow (0,0$)
transition by more than one order of magnitude and, hence, the line is very subthermally excited ($T_{\rm ex}=11$\,K). 
For the inferred abundance, the line is moderately optically thick ($\tau = 1$). 

The validity of the presented results rests on the assumption that the CO model by \cite{wilson01} is applicable 
to NH$_3$. The observation of the Orion Bar in the 
$(3,3)$ inversion line could provide a test, as this line is predicted to be the strongest NH$_3$ transition.


\begin{thebibliography}{}
        
        \bibitem[Cohen \& Poynter 1974]{cohenandpoynter74} Cohen E.A., Poynter R.L., 1974, J. Mol. Spec., 53, 131               

        \bibitem[Danby et al. 1988]{danby88} Danby G., Flower D.R., Valiron P., Schilke P., Walmsley C.M., 1988,
        		MNRAS, 235, 229

	\bibitem[Genzel \& Stutski 1989]{genzel89} Genzel R., Stutski J., 1989, ARAA, 27, 41
	
	\bibitem[Hermsen et al. 1988a]{hermsen88a} Hermsen W., Wilson T.L., Walmsley C.M., Henkel C., 1988a,
			A\&A, 201, 285	

	\bibitem[Hermsen et al. 1988b]{hermsen88b} Hermsen W., Wilson T.L., Bieging J.H., 1988b,
			A\&A, 201, 276		

	\bibitem[Ho et al. 1979]{ho79} Ho P.T.P., Barrett A.H., Myers P.C., et al., 1979, ApJ, 234, 912

	\bibitem[Ho \& Townes 1983]{hoandtownes83} Ho P.T.P., Townes C.H., 1983, ARAA, 21, 239
 
        \bibitem[Keene et al. 1983]{keene83} Keene J., Blake G.A., Phillips T.G., 1983, ApJ, 271, L\,27
        
        \bibitem[Liseau 2001]{liseau01} Liseau R., 2001, in: Pilbratt G.L., Cernicharo J., Heras A.M.,
        		Prusti T., and Harris R. (eds.), The Promise of the Herschel Space Observatory, 
        		ESA SP-460, p.\,313 

        \bibitem[Lovas 1992]{lovas} Lovas F.J., 1992, J. Phys. Chem. Ref. Data, 21, 181
        
        \bibitem[Morris et al. 1973]{morris73} Morris M., Zuckerman B., Palmer P., Turner B.E., 1973, ApJ, 186, 501
        
        \bibitem[Poynter \& Kakar 1975]{poynterandkakar75} Poynter R.L., Kakar R.K., 1975, ApJS, 29, 87
        
        \bibitem [Snell et al. 2000] {snell00} Snell R.L., Howe J.E., Ashby M.L.N., et al., 2000, ApJ, 539, L\,93       

        \bibitem[Tielens et al. 1993]{tielens93} Tielens A.G.G.M., Meixner M.M., van der Werf P.P., et al., Science 262, 86

        \bibitem[Townes \& Schawlow 1955]{townes1955} Townes C.H, Schawlow A.L., 1955, 
        		Microwave Spectroscopy, Dover Publications, Inc.

        \bibitem[Wilson et al. 2001]{wilson01} Wilson T.L., Muders D., Kramer C., Henkel C., 2001, ApJ, 557, 240

        \bibitem[Wiseman \& Ho 1998]{wisemanho98} Wiseman J.J., Ho P.T.P., 1998, ApJ, 502, 676
                
        \bibitem[van Dishoeck et al. 1993]{ewine93} van Dishoeck E.F., Blake G.A., Draine B.T., Lunine J.I., 1993,       		in: Levy E.H. \& Lunine J.I. (eds.), Protostars and Planets III, University of Arizona Press, p.\,163
\end{thebibliography}
\end{document}